\documentclass[9pt,twocolumn,twoside]{pnas-new}
\templatetype{pnasresearcharticle} 
\usepackage[english]{babel}
\usepackage{amsmath}
\usepackage{amsfonts}
\usepackage{mathtools}

\usepackage{graphicx}
\usepackage{hyperref}
\usepackage[label font={bf, normalsize}]{subcaption}     
\usepackage{color,soul}

\usepackage{dcolumn}
\usepackage{bm}

\usepackage{parskip}
\usepackage{booktabs}

\title{To what extent homophily and influencer networks explain song popularity}

\author[a]{Niklas Reisz}
\author[a]{Vito D.~P.~Servedio}
\author[abc]{Stefan Thurner\textsuperscript{1}}
\affil[a]{Complexity Science Hub Vienna, Josefst\"adter Strasse 39, A-1080 Vienna, Austria}
\affil[b]{Section for Science of Complex Systems, CeMSIIS, Medical University of Vienna, Spitalgasse 23, A-1090, Austria}
\affil[c]{Santa Fe Institute, 1399 Hyde Park Road, Santa Fe, NM 85701, USA}
\leadauthor{Niklas Reisz}

\significancestatement{
Machine learning models for the prediction of song popularity have become popular in the music industry. While acoustic features, genre, or artist popularity are known to drive popularity, less attention has been given to social aspects of music perception. Here we use the concept of homophily --the idea that social ties form predominantly between people with similar features--  to design a new predictive \emph{influence}-parameter. Using this parameter in machine learning models allows us to improve the prediction precision of whether a new song is likely to become popular by up to 50\%. We find that social influence is an essential factor in how listeners choose new songs and is at least as relevant as artist popularity or genre information. 
}

\authorcontributions{NR, ST, VDPS conceptualized the work and designed the model. NR collected and prepared the data and performed the analysis. NR and ST wrote the paper.}
\authordeclaration{The authors declare no conflict of interest.}
\correspondingauthor{\textsuperscript{1}To whom correspondence should be addressed. E-mail: stefan.thurner@meduniwien.ac.at }


\begin{abstract}
Forecasting the popularity of new songs has become a  standard practice in the music industry and provides a comparative  advantage for those that do it well. Considerable efforts were put into machine learning prediction models for that purpose. It is known that in these models, relevant predictive parameters include  intrinsic lyrical and acoustic characteristics, extrinsic factors (e.g., publisher influence and support), and the previous popularity of the artists. Much less attention was given to the social components of the spreading of song popularity. Recently, evidence for musical homophily --the tendency that people who are socially linked also share musical tastes-- was reported. Here we determine how musical homophily can be used to predict song popularity. The study is based on an extensive dataset from the \emph{last.fm} online music platform from which we can extract social links between listeners and their listening patterns. To quantify the importance of networks in the spreading of songs that eventually determines their popularity, we use musical homophily to design a predictive \textit{influence} parameter and show that its inclusion in state-of-the-art machine learning models enhances predictions of song popularity. The influence parameter improves the prediction precision (TP/(TP+FN)) by about 50\% from \boldsymbol{$0.14$} to \boldsymbol{$0.21$}, indicating that the social component in the spreading of music plays at least as significant a role as the artist's popularity or the impact of the genre.
\end{abstract}
\dates{\today}

\begin{document}

\maketitle
\ifthenelse{\boolean{shortarticle}}{\ifthenelse{\boolean{singlecolumn}}{\abscontentformatted}{\abscontent}}{}

\dropcap{W}hile you read this paper's abstract, more than 50 new songs were released worldwide. Global music production has long reached such high levels that it is no longer possible to listen to every new song. On the world's largest music streaming platform {\em Spotify}, roughly 136 days worth of music are published daily \cite{spotifypost, spotifypost2}. This means that more music is produced in a year than one could listen to in an entire lifetime. In this ocean of new songs, there is a growing need for efficient selection and filtering, and the markets for attention have become increasingly competitive. This becomes apparent in increasingly imbalanced distributions of song popularity. While the majority of songs receive little to no attention, a small fraction become hits that are listened to billions of times \cite{Hu_2008}. 
Predicting which songs have the potential to become one of these hits has become an increasingly important task for music publishers \cite{rosen_1981}.  The issue has sparked  substantial commercial interest as publishers focus their marketing activities on high-potential songs and artists \cite{Pham_2016}. 
For quantitative predictions, a multitude of different approaches has been explored. The vast volume of data made available by streaming platforms in the past decade triggered a boost of data-driven approaches.

In the early 2000s, a discussion started on the possibility of quantitatively predicting song popularity. Using traditional machine learning classifiers on lyrical and acoustic song attributes, it was attempted to identify hits prospectively \cite{Dhanaraj_2005}.
It was concluded that predictions were significantly better than random, with the lyrical features slightly outperforming the  acoustic ones.
In \cite{Pachet_2008}, it was claimed that the science of predicting hits, the so-called \textit{hit song science}, was not yet a science as they demonstrated that the usual machine learning methods were not able to forecast the success of songs. This assertion was soon challenged in \cite{Ni_2011}, where the authors argued that, given a relevant set of acoustic song features, forecasting song popularity was possible with more specific machine learning approaches. In these early works, the emphasis was primarily on \textit{intrinsic} acoustic and lyrics features. Later, other studies improved outcomes with more sophisticated methods and expansive datasets. Specific song attributes such as ``happiness'', ``partyness'' and repetitiveness were found to increase the chances of songs becoming hits \cite{Interiano_2018, Lassche_2019}. Deep convolutional structures and machine learning regressions were convincingly shown to have predictive power \cite{Yang_2017, Middlebrook_2019, MartinGutierrez2020}.

While these studies focused exclusively on \textit{intrinsic} properties of songs, newer studies have included \textit{extrinsic} features such as metadata or artist popularity. It was claimed that extrinsic factors often carry increased predicting power, the most predictive feature being the previous popularity of the artist \cite{Pham_2016}. ~\cite{Askin_2017, Shin_2018} confirmed this result by finding that both, the previous popularity of the artist and the support of the publisher, have a significant impact on the success chances of songs. In addition to having increased chances of making it into the charts, \cite{Hyunsuk_2018} found that songs of well-known artists and big publishers also stay there longer. Reference~\cite{Kim_2021} showed that superstars significantly dominate the market share and measured a positive correlation between song success and the number of songs an artist has released previously.
For \emph{extrinsic} song attributes, \cite{Yu_2019} found that Support Vector Regression (SVR) performs slightly better than neural networks.

Predictive indicators were also found in Twitter data  \cite{Kim_2014} where music-related hashtags such as \emph{\#nowplaying} were counted. The number of daily tweets about specific songs and artists is highly correlated with the listening trend of that song. Counting the same tags, \cite{Tsiara_2020} found a moderate correlation between the number of tweets and the chart position of a song, as well as a weak correlation with the sentiment of the tweets.
These Twitter-based correlations hint at a social component to the success of songs and that this information might be encoded in social networks. A similar thought was followed in \cite{Rosati_2021}, where the spread of song popularity was modeled with a SIR disease spreading model. There, it is argued that social processes that lead to the spread of music are similar to those of disease spreading. For some genres, for which social connectivity is higher, songs appear to spread faster.

The fact that social networks co-create homophily --the tendency that people that share common treats are more likely to form social ties-- has been observed in a variety of contexts, ranging from obesity \cite{Nicholas_Christakis_obesity} to performance in schools \cite{smirnovthurner}. Also, in music listening behavior, the concept of homophily has been studied \cite{McPherson_2001}. 
Recently, homophily's importance in relation to music preference was confirmed in a study on 1144 early adolescents, where music preference plays a significant role in selecting friends \cite{Franken2017}.
The 
online music-listening platform and social network \emph{last.fm}\footnote{\url{https://www.last.fm/}} offers an excellent ground for studying homophily as it simultaneously provides access to both social links and music listening records of users \cite{doi:10.1504/IJWBC.2011.039513}.
In \cite{10.1145/2380718.2380725}, the authors study \emph{last.fm} data to predict friendship links. They find that music preference alone rarely leads to friendships. In most cases, sharing friends is the best predictor of future friendships, i.e. triadic closure that has been predicted in \cite{granovetter} and explains basic structures of social multilayer networks \cite{klimekthurner}. 
Reference~\cite{Guidotti_2020} confirmed that people with similar music preferences tend to cluster, indicating that friends tend to listen to similar music.
Homophily in music listening was found in explorative behavior \cite{Duricic_2021}. By estimating how mainstream, novel, or diverse listening records of users are, they find that highly explorative people tend to look for friends that are similar.
Reference \cite{DiBona_2022} confirmed that result and designed a model that explains the finding, stressing that highly explorative users tend to be friends with users with similar discovery rates.
While \cite{Duricic_2021} use homophily to predict friendship links on \emph{last.fm}, in \cite{P_lovics_2015, Palovics_2013} it is investigated how information about new artists spreads through social links and quantified to which extent user behavior is copied. It is demonstrated how homophily can be leveraged to improve song recommendations by recommending new songs to users shortly after close friends are listening to them.


While acoustic and metadata-based success factors for predicting song popularity have been studied extensively, much less attention has been given to the influence of social interactions. It has been shown that these influence listening behavior significantly. In this work, we quantify how social interactions, in particular homophily, impact song popularity. We demonstrate how homophily emerges as users influence their friends to listen to specific songs, which drives the discovery and popularity of new songs through cascades of song recommendations. We quantify this user influence by state-of-the-art machine learning predictions of song popularity by estimating to what extent its predictions can be improved.

{\em Homophily.}
To estimate homophily, we  derived a dataset from the online music-listening platform and social network {\em last.fm}. Our data includes 300 million individual listening events of 10 million songs. It is enriched by the (undirected and unweighted) friendship network, where nodes represent users and links represent bidirectional friendships. It consists of 2.7 million nodes and 15 million links with an average degree of 11.6. The network is connected with a diameter of 6. This dataset allows us to link friendships to music-listening histories. For more details, see SI text~\ref{SI1}.

We first determine the music preferences of every user. Music preferences can be assessed in {\em last.fm} through user-defined tags. Users define and add tags such as \textit{Rock}, \textit{80s}, or \textit{Acoustic} to songs, artists, or albums. Here, we focus on artist tags because they are more abundant than song or album tags and offer more reliable statistics. Every artist, $a$, can have multiple tags, $t$, which are represented as weights, $w_{at}$, ranging between 1 and 100, based on their frequency of appearance (100 corresponds to the most frequent tag for that artist). We then compute a music preference matrix, $m_{ut}$, by summing over the weights, $w_{at}$, for each tag, $t$, of each artist, $a$, for each time a user, $u$, is listening to a song by that artist. The music preference matrix, $M$, with elements $m_{u, t}$ is hence defined as
\begin{equation}
    m_{ut} = \sum_{a, s} l_{us} i_{sa} w_{at} \, ,
\end{equation}
where $l_{us}$ is the number of times user, $u$, listened to song, $s$. $i_{sa}=1$ if artist, $a$, interpreted song, $s$, and is zero otherwise. In the music preference matrix $M$, each row corresponds to a vector of music preference of one user, expressed by the weighted artist tags. To compare the music preference vectors of users $i$ and $j$, we use the cosine-similarity, defined as
\begin{equation}
    \cos(\theta_{ij}) = \frac{\sum_t m_{u=i, t} m_{u=j, t}}{\sqrt{\sum_t m_{u=i, t}^2} \sqrt{\sum_t m_{u=j, t}^2}} \, .
\end{equation}
The idea will be to compare the alignment of music preference vectors for users that are friends to the alignment between randomly picked users (that are very likely not friends). 

\begin{figure*}[htb]

  \begin{subfigure}[b]{0.5\linewidth}  
    \includegraphics[width=\linewidth]{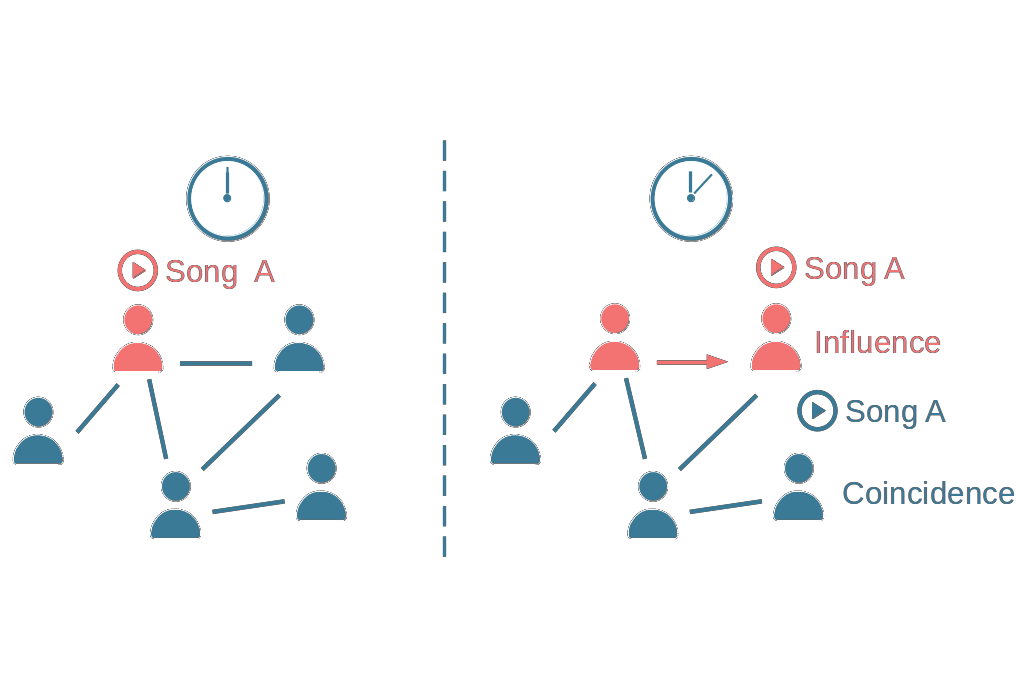}
    \caption{} 
    \label{fig:stasdet0}   
  \end{subfigure}
  \hfill 
  \begin{subfigure}[b]{0.475\linewidth}
    \includegraphics[width=\linewidth]{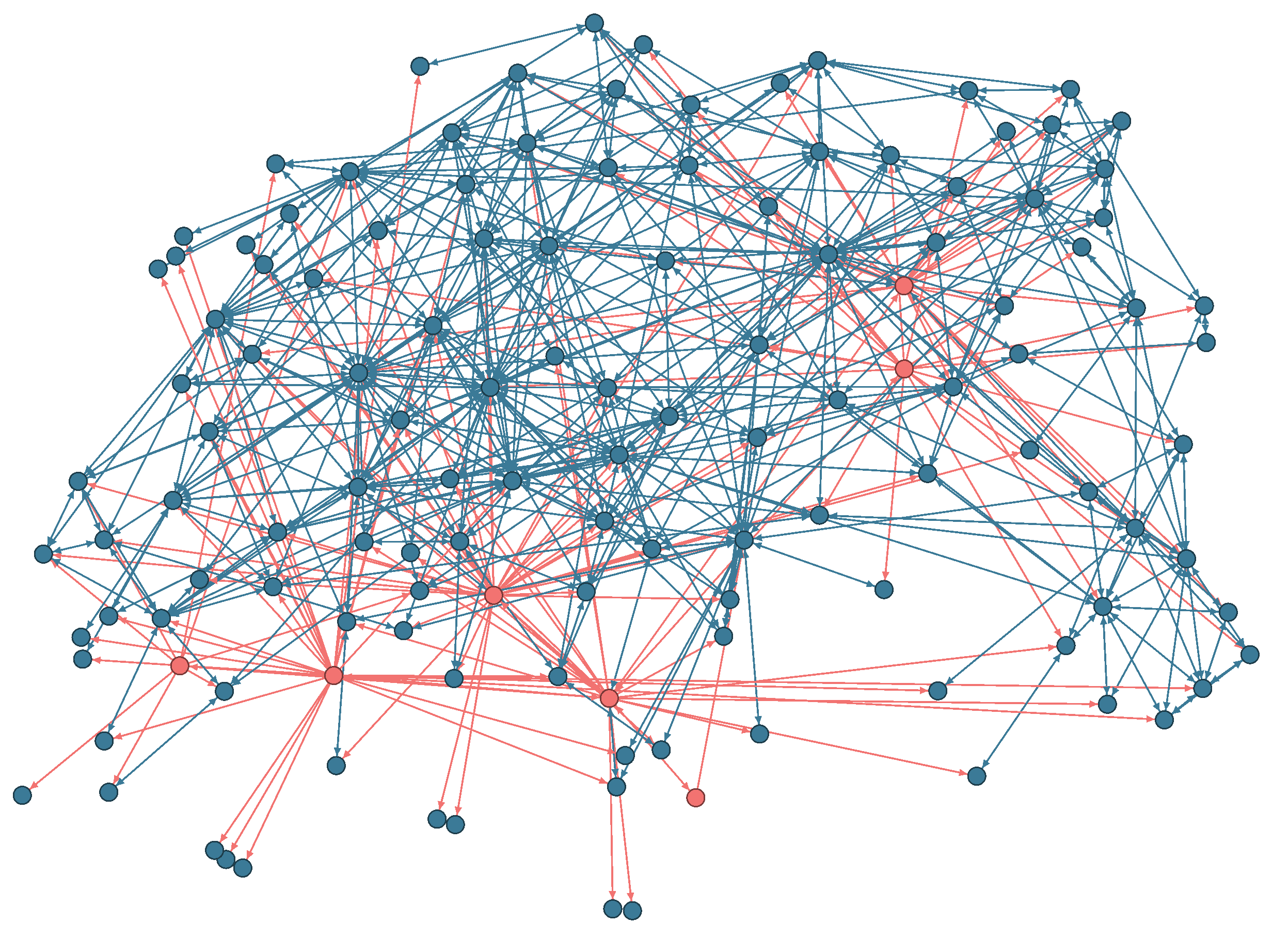} 
    \caption{}
    \label{fig:stasdet1}   
  \end{subfigure} 

  \bigskip 
  \begin{subfigure}[b]{0.95\linewidth}
    \includegraphics[width=\linewidth]{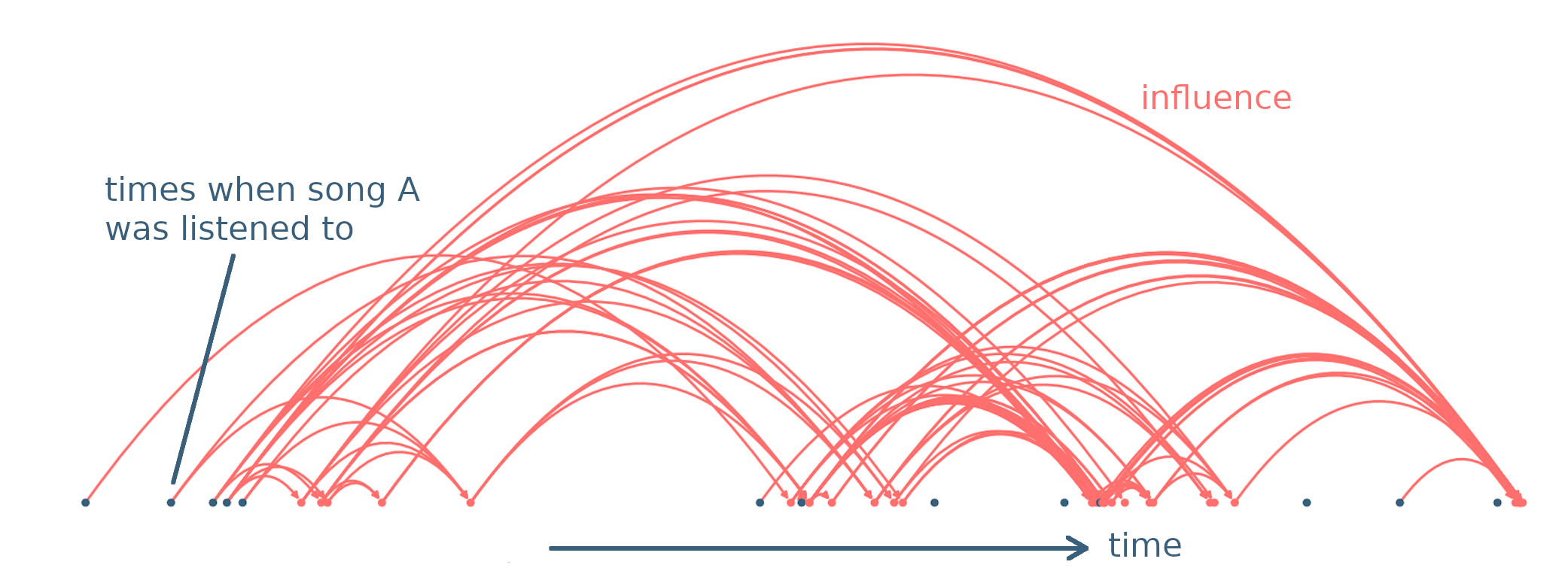} 
    \caption{}
    \label{fig:stasdet2}
  \end{subfigure}
  
  \caption{
    (a) Schematic depiction of user influence. If a user listens to a song at time $t_1$ and a friend of that user listens to the same song for the first time at time $t_2$ with $t_2 - t_1< \Delta$ smaller than some threshold $\Delta$, the second user is said to have been influenced by the first with respect to that song. The influencer network is shown in orange, the friendship network in blue. The influencer network is a sub-network of the friendship network. 
    (b) A fraction of the influencer network of {\em last.fm} from a 24h time window. Several strong influencers are visible (influencers and influencing links in orange) within the friendship network (blue).
    (c) Example of a timeline of a single song. Dots along the x-axis represent different users that listen to a song for the first time (location of dot). Dots are ordered in time, from left to right. Blue dots are users that found the song on their own, while orange dots represent users that were influenced by their friends. Arrows mark influencing events, where one user influences another into listening to a song.}
  \label{fig1}
\end{figure*}

{\em Influencer network.} As a next step, we define the influencer network. Influence is the number of times a user's friends copied the listening behavior of that user within a specified time window. We define it algorithmically. Whenever a friend, $u_j$, of a user, $u_i$, listens to a song for the first time at time $t_j$ shortly after user, $u_i$, listened to that song at time $t_i$, the influence, $I_{ij}$, of user, $i$, on user, $j$, increases by one if  $t_j - t_i< \Delta$ is smaller than some threshold $\Delta$. A sketch of the concept and the derived influencer network are shown in Fig.~\ref{fig1}. The influencer network (orange) is a directed network where users are nodes, and a weighted link represents the strength of a user's influence over another. In our case, it consists of 9200 nodes and 190,000 links, with an average degree of 21. The influencer network is connected and has a diameter of 11. Since friendship is a requirement for influence, the influencer network is a sub-network of the friendship network (blue), all nodes and links present in the influencer network also exist in the friendship network. Influence on {\em last.fm} is possible because users can see which songs their friends are listening to and can give specific song recommendations to them. 

{\em Prediction strategy.} 
To simplify the prediction task of the future success of a song, we only classify songs into average songs and future hit songs. We define hit songs as songs with at least 1000 listenings in any one month, which corresponds to the top 1\% of songs. We do this classification based on an initial sample of features or parameters derived from the first 200 times a new song has been listened to. We define three  classes of predictive parameters: preferential-attachment-based, time-based, and homophily-based. The preferential-attachment-based parameters are (i) the previous popularity of the artist and (ii) the genre's popularity. The time-based parameters are (iii) the time needed to reach 200 listenings, (iv) the tendency of users to re-listen to a song, and (v) the temporal trend quantified by the area under the curve of the cumulative listening count as a function of time. We also include three variations of these parameters, (vi) the number of users that listen to the song at least twice within a week, (vii) a normalized variation of (v), where we take the average y-value instead of the area, and (viii) a variation of (v) where we subtract the y-value from the diagonal and then compute the area. These constitute the basic eight parameters of the model. In addition to these parameters, we define homophily-based parameters and compute them for every song. For these, we identify the users that contributed to the first 200 times a song has been listened to. We use (ix) the influence scores, as defined above, as well as (xii) the cosine similarity between users and their friends as homophily-based parameters. Finally, we compute (xiii) the degree, (xiv) the PageRank,  (xv) the nearest neighbor degree, and (xvi) the clustering coefficient both on the friendship network and the influencer network (xvii-xxi). All parameters are described in detail in SI text~\ref{SI:predictionparams}. We use these parameters as input for a machine learning ensemble, see methods.


\section*{Results}

\begin{figure}[tb]
\centering
\includegraphics[width=1\columnwidth]{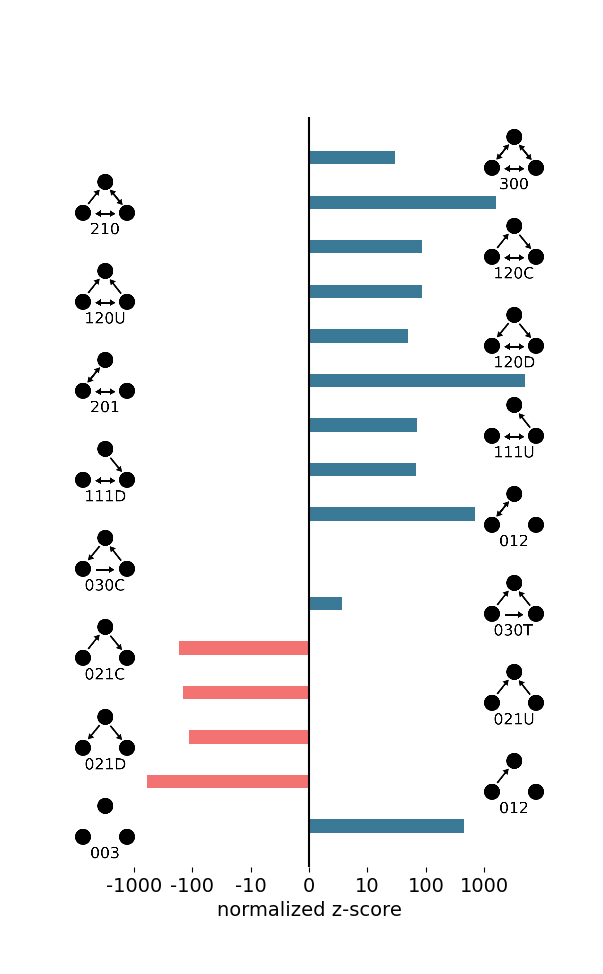}
\caption{Triadic census of the influencer network. Comparison to a shuffled network (configuration model), averaged over 100 random shufflings. Triad names follow the convention of \cite{triads}.}
\label{fig:triadsif}
\end{figure}

{\em Influencer network.}
Analyzing the triad statistics of the influencer networks,  we find high levels of reciprocity in the influencer network, see Fig. \ref{fig:triadsif}. This is seen by the fact that triangles that include reciprocal links are strongly over-represented (blue bars) with respect to a random graph  with the same number of nodes and (directed) links (shuffled network, configuration model). Triangles with no reciprocity are under-represented (red). This finding indicates that, generally, influence goes in both directions. Users who discover a new song from a friend (getting influenced) often influence other friends into listening to the song, leading to cascades, such as shown in Fig.~\ref{fig1} (c).
In the SI text~\ref{SI:influ}, we identify the ``influencers'' and ``followers'' by comparing the number of times users got influenced with how often they influenced others.

\begin{figure}[tb]
    \centering
    \includegraphics[width=0.95\columnwidth]{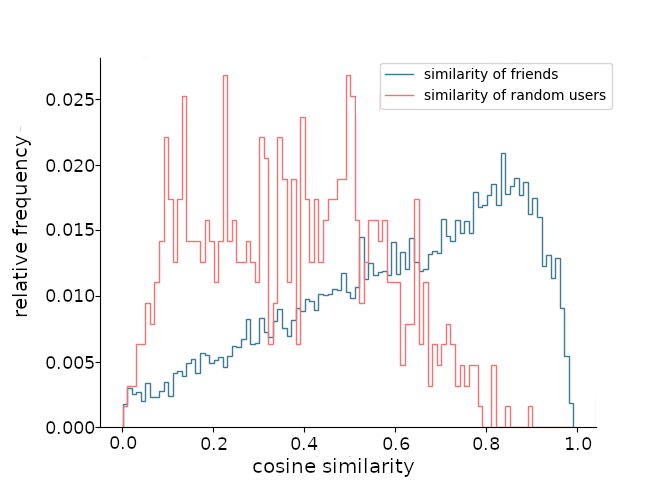}
    \caption{Cosine similarity distribution for users and their friends (blue), and between random users (red) that are typically not befriended. From the distribution, it becomes apparent that very similar people are almost certainly friends on {\em last.fm.}
	}
	\label{fig:cosinesim}
\end{figure}

{\em Homophily results.}
We confirm the presence of strong homophily among users that are friends on {\em last.fm} by comparing the musical preferences of 1000 randomly picked users to those of their friends. We find an average cosine-similarity of musical-preference vectors $\Vec{m}_{u}$ of $ \cos(\theta) = 0.58$. In contrast, when comparing the musical preference vectors to an equal number of randomly picked users, the average cosine-similarity drops 
to $\cos(\theta) = 0.25$. A histogram of the respective cosine-similarities is shown in Fig.~\ref{fig:cosinesim}. The $t$-test for independent samples has a test statistic of 32 and a $p$-value of $p \le 10^{-200}$. If the entries of the music preference vectors of the randomly picked users are shuffled in a random fashion, similarity decreases to levels below $10^{-4}$ and essentially disappears. 

We find that both the user's influence and the tendency to get influenced correlate with the average similarity in musical preferences. When comparing users to their friends, this correlation is highly significant with a $p \leq 10^{-4}$. In contrast, when comparing random users, we do not find any correlation. For more details, see SI text~\ref{SI:hom_vs_inf}.

{\em Improving predictions.}
Both the homophily score and the influence score correlate with song popularity. Parameters that correlate with popularity can be tested if they also predict it. The Pearson correlation coefficients, $r$, between all parameters and the song popularity are found in Tab. \ref{tab1}. The strongest (anti) correlation for song popularity we find for the area under the temporal listening trend with $r=-0.22$. Other time-based parameters have comparable correlations. The influence score correlates better ($r=0.17$) with song popularity than the previous popularity of the artist  ($r=0.14$). Also, influencer network based parameters correlate similarly as artist popularity.

\begin{center}
\begin{table}[tb]
\centering
\caption{Pearson correlation coefficients between prediction parameters and song popularity. Bold numbers show the strongest correlations for homophily-based and baseline parameters. The prediction parameters were computed on the first 200 listenings for each song. Song popularity was defined as the maximum number of times a song was listened to {\em last.fm} in its best month. All p-values are below \boldsymbol{$10^{-200}$}. Only the best-performing variations of parameters are listed here. All other variations can be found in SI text~\ref{SI:predictionparams}. 
}
\begin{tabular}{ l c }
parameter/feature & r \\
\hline
degree (friendship network) & 0.090 \\
degree (influencer network) &  0.132\\
PageRank (influencer network) & 0.110 \\
influence & \textbf{0.174} \\
homophily & 0.081 \\
 & \\
time to reach 200 listenings & -0.201 \\
genre average & 0.069 \\
time between repeated listenings & -0.193 \\
area under the trend curve & \textbf{-0.223} \\
previous popularity of artist & 0.135 \\
\end{tabular}
\label{tab1}
\end{table}
\end{center}

Table \ref{tab2} shows the prediction results for three different models quantified by accuracy, precision, and recall. We predict if a song becomes a hit-song or an average song based on information from the 200 first listenings and the structure of the influencer and friendship networks. 
A hit song is defined as a song that is listened to at least 1000 times in its best month, putting it in the top 1\% of all songs. For the classification task, we use a machine learning ensemble including classifiers based on Support Vector Classification, Random Forest, Ada Boost, Gradient Boost, K-Neighbors, and a multilayer perceptron neural network, see Methods. Based on the input parameters, the machine learning models try to classify each song into one of two classes: hit songs or regular songs.
In the first model, which we refer to as the combined model (a), we use the combined parameter sets with preferential-attachment-, time-, and homophily-based parameters. In the second, the social networks only scenario (b), we use the homophily-based parameters only. In the third, we combine the preferential-attachment and time-based parameters without using the homophily-based parameters and refer to it as the baseline model (c). The baseline model is based on commonly used parameters from the literature and forms the baseline against which we want to compare the results of models (a) and (b). 

The strongest result is the improvement of hit-precision by 50\% in the combined model (a) when compared to the baseline model (c). In addition, non-hit recall and accuracy improved by one percentage point at an already  high level. On the downside, hit-recall decreased by around 14\%. Adding the homophily-based influence score as a predictive parameter significantly improves accuracy and precision, as well as non-hit recall, at the cost of hit-recall. This suggests that the combined model (a) is able to highly reduce the number of false positives at the cost of missing a small number of true positives.
The homophily-based model (b) on its own manages to already identify 40\% of all hit songs correctly and reaches an accuracy of 95\%. We see that while conventional models like our baseline model (a) are superior to the homophily-based model (b), the homophily-based model performs already on a high level, and a combination of both leads to the best results.

\begin{center}
\begin{table}[tb]
\caption{Classification results for three different cases: (a) the combined model (preferential attachment, time, and homophily), (b) the model with only homophily-based social network parameters, and (c) the baseline model without explicit social network information.
We see that a combined model (a) performs best, with the highest accuracy, hit precision, and non-hit recall. The homophily-based model (b) performs well on its own but is outclassed by the baseline model. 
}
\begin{tabular}{ l l c }

model & prediction category &  \\
\hline
(a) combined model    & accuracy &         \textbf{0.98}   \\
    & hit precision &     \textbf{0.21}  \\
    & non-hit precision & 1.0   \\
    & hit recall &        \textbf{0.60}  \\
    & non-hit recall &    \textbf{0.99}   \\
\hline
(b) social network only    & accuracy &         0.95   \\
    & hit precision &     0.05   \\
    & non-hit precision & 1.0   \\
    & hit recall &        0.40  \\
    & non-hit recall &    0.96   \\
\hline
(c) without social network    & accuracy &         \textbf{0.97}   \\
    & hit precision &     \textbf{0.14}   \\
    & non-hit precision & 1.0   \\
    & hit recall &        \textbf{0.70}   \\
    & non-hit recall &    \textbf{0.98}  \\

\end{tabular}
\label{tab2}
\end{table}
\end{center}

\section*{Discussion}

We demonstrated how social interactions can be used to enhance song popularity predictions using a large dataset collected from the online platform {\em last.fm}. From an influencer network we derive an influence score for every user that captures their tendency to influence others to imitate their listening behavior. Based on a small sample of first listenings, we compute several metrics on the influencer network and use these as the predictive parameters in a machine learning classifier ensemble to categorize songs into potential hits and average songs. Our model exhibits up to 50\% improved precision over a baseline model that uses common preferential attachment and time-based parameters.

The used concept of influencing is closely related to homophily. Using the music preference vectors we estimate the similarity of tastes and find that people that are related through friendship links tend to have aligned tatses -- i.e. strong homophily is confirmed, consistent with earlier findings of \cite{Guidotti_2020, P_lovics_2015, Zhou_2018, Bisgin_2011}. Influence represents the degree to which users can make their friends listening to the same music. This increases the similarity in the music preference vectors and drives homophily. We show that while some people tend to pioneer new music tastes and others tend to follow these pioneers, in most cases, influencer interactions go in both directions. This means that users that get influenced by their friends can in turn influence multiple other friends, leading to network structures that enable cascading spreading of new songs, fueling the new song's popularity.

Influence between users contributes to preferential attachment. The more people listen to a song, the more likely it gets recommended to friends, thus the probability of it being listened to increases.
Parameters that relate to either preferential attachment, such as the previous popularity of the artist, or to forgetting, such as the listening trend, have been widely used in previous works to predict the popularity of songs \cite{Pham_2016, Askin_2017, Shin_2018, Hyunsuk_2018}.
In this study, we focus on these extrinsic properties and contribute new parameters that have the potential to improve the performance of existing models. 

There are several severe limitations. Obviously, {\em last.fm} only provides a partial view on what is going on in music listening and recommendations. There are many other channels where people listen to music and exchange information about new songs which leads to a ``cold-start'' problem: a song that is new on the {\em last.fm} platform doesn't need to be new to its users. Hence, song popularity predictions might get offset by events that are beyond the scope of the available data.
This is in part related also to the issue of comparability. A multitude of different datasets is used across the literature, each of them limited in some aspect, for instance, to a specific platform or a geographic region \cite{Interiano_2018, Yang_2017, Middlebrook_2019, Pham_2016}. This is coupled with an apparent lack of a consistent definition of hit songs. In addition to that, in models with many parameters and hyperparameters, performance might be optimized with regard to different metrics. Altogether, specific results are difficult to compare across different studies. For this reason, we chose to use our own baseline model as a benchmark and aim for precise and intuitive definitions of popularity and hit songs, such that different machine learning models might be compared in a consistent way.
Further, to some degree, popularity predictions might be self-fulfilling prophecies. It has been observed that people tend to reproduce perceived song popularity that is presented to them, even if these have been heavily modified \cite{Salganik2006, Salganik2008, Lynn2016}. 
Finally, there is also an ongoing discussion on whether artists should focus on improving popularity over other goals. Most popular doesn't necessarily imply most enjoyable or most relevant \cite{Monechi_2017}. Rather, it indicates high commercial relevance, which might not necessarily be the top priority.
However, even given these shortcomings, in this work, we were able to compare the predictions within a closed framework. Our main result is that we are able to find a substantial relative increase in predictability by including social information. The concept presented here can straightforwardly be transferred to other domains such as movies, books, posts, or even physical goods. Given the growing availability of social network data, comparable approaches might further uncover the social underpinnings of consumer behavior in our society.

\section*{Methods}
\label{sec:methods}

For song popularity predictions, we use a machine learning ensemble. The ensemble includes classifiers based on Support Vector Classification, Random Forest, Ada Boost, Gradient Boost, K-Neighbors, and a neural network. The different classifiers hold a majority vote on the classification of each song. Each song is classified as either a future hit-song or an average song, where hit-songs are defined as songs in the top 1\% of songs, which equates to being listened to at least 1000 times in the best month in our dataset. In the following, we give a brief overview of the machine learning models used, the specific implementation, and their hyperparameter settings.

Support Vector Classification is a supervised learning model that tries to map training data into a higher dimensional space with the aim of maximizing the gap between points of different classes in that space \cite{svm}. In this study, we use the Python sklearn implementation \cite{svmpy} with balanced class weights and a value of $C=1$ for the regularization hyperparameter.

Random Forest classification is a supervised learning model that builds multiple randomized decision trees based on the training set \cite{Biau2016}. The classification outcome is then the majority vote of the individual trees. In this study, we use the Python sklearn implementation \cite{rfpy} with 100 estimators and balanced class weights.

Ada Boost, short for adaptive boosting, is a classifier that iteratively learns from the mistakes of weak classifiers, turning them into strong classifiers \cite{Schapire2013}. In this study, we use the Python sklearn implementation \cite{adbpy} with 100 estimators.

Gradient Boost classifier is a classifier that works as an ensemble of weak classifiers, typically decision trees. These weak classifiers are gradually added during the learning process while aiming for maximum correlation with the negative gradient of the loss function \cite{natekin2013gradient}. In this study, we use the Python sklearn implementation \cite{gbpy} with 100 estimators, a learning rate of 1, and a maximum depth of 1.

K-nearest neighbor classification is based on a multidimensional feature space that is populated by the feature vectors of the training set and their labels \cite{knn}. During classification, one looks at the nearest k neighbors of an element and attaches the label to it that is most common among its neighbors. In this study, we use the Python sklearn implementation \cite{knnpy} with the number of nearest neighbors k equal to 5.

A multilayer perceptron is a fully connected, feed-forward artificial neural network that consists of at least three layers: an input, a hidden layer, and an output layer \cite{Hastie2009}. All nodes in each layer are connected to all other nodes of the following layer. It uses non-linear activation functions and learns through back-propagation. In this study, we use the Python sklearn implementation \cite{mlppy} with a maximum number of iterations of 1000.

These machine learning models are each trained on (the same) 60\% of the data and then used to classify the other 40\%. The data is classified according to the majority vote of the different models. 

\renewcommand\bibname{References}
\section*{References}
\bibliography{library}

\begin{thebibliography}{10}

\bibitem{spotifypost}
T Ingham, Over 60,000 tracks are now uploaded to spotify every day. that’s
  nearly one per second
  (\url{https://www.musicbusinessworldwide.com/over-60000-tracks-are-now-uploaded-to-spotify-daily-thats-nearly-one-per-second/})
  (2021) Accessed: 2022-08-25.

\bibitem{spotifypost2}
B Houghton, 60,000 tracks are uploaded to spotify every day
  (\url{https://www.hypebot.com/hypebot/2021/02/60000-tracks-are-uploaded-to-spotify-every-day.html})
  (2021) Accessed: 2022-08-25.

\bibitem{Hu_2008}
HB Hu, DY Han, Empirical analysis of individual popularity and activity on an
  online music service system.
\newblock {\em\protect\JournalTitle{Physica A: Statistical Mechanics and its
  Applications}} \textbf{387}, 5916--5921 (2008).

\bibitem{rosen_1981}
S Rosen, The economics of superstars.
\newblock {\em\protect\JournalTitle{The American economic review}} \textbf{71},
  845--858 (1981).

\bibitem{Pham_2016}
J Pham, E Kyauk, E Park, Predicting song popularity.
\newblock {\em\protect\JournalTitle{Dept. Comput. Sci., Stanford Univ.,
  Stanford, CA, USA, Tech. Rep}} \textbf{26} (2016).

\bibitem{Dhanaraj_2005}
R Dhanaraj, B Logan, Automatic prediction of hit songs in {\em proceedings of
  the international conference on music information retrieval (ISMIR)}.
\newblock pp. 488--491 (2005).

\bibitem{Pachet_2008}
F Pachet, P Roy, Hit song science is not yet a science. in {\em ISMIR}.
\newblock pp. 355--360 (2008).

\bibitem{Ni_2011}
Y Ni, R Santos-Rodriguez, M Mcvicar, T De~Bie, Hit song science once again a
  science in {\em 4th International Workshop on Machine Learning and Music}.
\newblock (Citeseer), (2011).

\bibitem{Interiano_2018}
M Interiano, et~al., Musical trends and predictability of success in
  contemporary songs in and out of the top charts.
\newblock {\em\protect\JournalTitle{Royal Society Open Science}} \textbf{5},
  171274 (2018).

\bibitem{Lassche_2019}
A Lassche, F Karsdorp, E Stronks, Repetition and popularity in early modern
  songs in {\em DH 2019: Proceedings of the 2019 Digital Humanities
  Conference}.
\newblock (2019 Digital Humanities Conference), (2019).

\bibitem{Yang_2017}
LC Yang, SY Chou, JY Liu, YH Yang, YA Chen, Revisiting the problem of
  audio-based hit song prediction using convolutional neural networks in {\em
  2017 IEEE International Conference on Acoustics, Speech and Signal Processing
  (ICASSP)}.
\newblock pp. 621--625 (2017).

\bibitem{Middlebrook_2019}
K Middlebrook, K Sheik, Song hit prediction: Predicting billboard hits using
  spotify data.
\newblock {\em\protect\JournalTitle{CoRR}} \textbf{abs/1908.08609} (2019).

\bibitem{MartinGutierrez2020}
D Martin-Gutierrez, GH Penaloza, A Belmonte-Hernandez, FA Garcia, A multimodal
  end-to-end deep learning architecture for music popularity prediction.
\newblock {\em\protect\JournalTitle{{IEEE} Access}} \textbf{8}, 39361--39374
  (2020).

\bibitem{Askin_2017}
N Askin, M Mauskapf, What makes popular culture popular? product features and
  optimal differentiation in music.
\newblock {\em\protect\JournalTitle{American Sociological Review}} \textbf{82},
  910--944 (2017).

\bibitem{Shin_2018}
S SHIN, J PARK, On-chart success dynamics of popular songs.
\newblock {\em\protect\JournalTitle{Advances in Complex Systems}} \textbf{21},
  1850008 (2018).

\bibitem{Hyunsuk_2018}
H Im, H Song, J Jung, A survival analysis of songs on digital music platform.
\newblock {\em\protect\JournalTitle{Telematics and Informatics}} \textbf{35},
  1675--1686 (2018).

\bibitem{Kim_2021}
ST Kim, JH Oh, Music intelligence: Granular data and prediction of top ten hit
  songs.
\newblock {\em\protect\JournalTitle{Decision Support Systems}} \textbf{145},
  113535 (2021).

\bibitem{Yu_2019}
H Yu, Y Li, S Zhang, C Liang, Popularity prediction for artists based on user
  songs dataset in {\em Proceedings of the 2019 5th International Conference on
  Computing and Artificial Intelligence}, ICCAI '19.
\newblock (Association for Computing Machinery, New York, NY, USA), p. 17–24
  (2019).

\bibitem{Kim_2014}
Y Kim, B Suh, K Lee, \#nowplaying the future billboard: Mining music listening
  behaviors of twitter users for hit song prediction in {\em Proceedings of the
  First International Workshop on Social Media Retrieval and Analysis}, SoMeRA
  '14.
\newblock (Association for Computing Machinery, New York, NY, USA), p. 51–56
  (2014).

\bibitem{Tsiara_2020}
E Tsiara, C Tjortjis, Using twitter to predict chart position for songs in {\em
  Artificial Intelligence Applications and Innovations}, eds.{} I Maglogiannis,
  L Iliadis, E Pimenidis.
\newblock (Springer International Publishing, Cham), pp. 62--72 (2020).

\bibitem{Rosati_2021}
DP Rosati, MH Woolhouse, BM Bolker, DJD Earn, Modelling song popularity as a
  contagious process.
\newblock {\em\protect\JournalTitle{Proceedings of the Royal Society A:
  Mathematical, Physical and Engineering Sciences}} \textbf{477}, 20210457
  (2021).

\bibitem{Nicholas_Christakis_obesity}
NA Christakis, JH Fowler, The spread of obesity in a large social network over
  32 years.
\newblock {\em\protect\JournalTitle{New England Journal of Medicine}}
  \textbf{357}, 370--379 (2007).

\bibitem{smirnovthurner}
I Smirnov, S Thurner, Formation of homophily in academic performance: Students
  change their friends rather than performance.
\newblock {\em\protect\JournalTitle{{PLOS} {ONE}}} \textbf{12}, e0183473
  (2017).

\bibitem{McPherson_2001}
LSL Miller~McPherson, JM Cook, Birds of a feather: Homophily in social
  networks.
\newblock {\em\protect\JournalTitle{Annual Review of Sociology}} \textbf{27},
  415--444 (2001).

\bibitem{Franken2017}
A Franken, L Keijsers, JK Dijkstra, T ter Bogt, Music preferences, friendship,
  and externalizing behavior in early adolescence: A {SIENA} examination of the
  music marker theory using the {SNARE} study.
\newblock {\em\protect\JournalTitle{Journal of Youth and Adolescence}}
  \textbf{46}, 1839--1850 (2017).

\bibitem{doi:10.1504/IJWBC.2011.039513}
P Mechant, T Evens, Interaction in web-based communities: a case study of
  last.fm.
\newblock {\em\protect\JournalTitle{International Journal of Web Based
  Communities}} \textbf{7}, 234--249 (2011).

\bibitem{10.1145/2380718.2380725}
K Bischoff, We love rock 'n' roll: Analyzing and predicting friendship links in
  last.fm in {\em Proceedings of the 4th Annual ACM Web Science Conference},
  WebSci '12.
\newblock (Association for Computing Machinery, New York, NY, USA), p. 47–56
  (2012).

\bibitem{granovetter}
MS Granovetter, The strength of weak ties.
\newblock {\em\protect\JournalTitle{American journal of sociology}}
  \textbf{78}, 1360--1380 (1973).

\bibitem{klimekthurner}
M Sadilek, P Klimek, S Thurner, Asocial balance{\textemdash}how your friends
  determine your enemies: understanding the co-evolution of friendship and
  enmity interactions in a virtual world.
\newblock {\em\protect\JournalTitle{Journal of Computational Social Science}}
  \textbf{1}, 227--239 (2017).

\bibitem{Guidotti_2020}
R Guidotti, G Rossetti, ``know thyself'' how personal music tastes shape the
  last.fm online social network in {\em Formal Methods. FM 2019 International
  Workshops}, eds.{} E Sekerinski, et~al.
\newblock (Springer International Publishing, Cham), pp. 146--161 (2020).

\bibitem{Duricic_2021}
T Duricic, D Kowald, M Schedl, E Lex, My friends also prefer diverse music:
  Homophily and link prediction with user preferences for mainstream, novelty,
  and diversity in music in {\em Proceedings of the 2021 IEEE/ACM International
  Conference on Advances in Social Networks Analysis and Mining}, ASONAM '21.
\newblock (Association for Computing Machinery, New York, NY, USA), p.
  447–454 (2022).

\bibitem{DiBona_2022}
G Di~Bona, et~al., Social interactions affect discovery processes (2022).

\bibitem{P_lovics_2015}
R P{\'{a}}lovics, AA Bencz{\'{u}}r, Temporal influence over the last.fm social
  network.
\newblock {\em\protect\JournalTitle{Social Network Analysis and Mining}}
  \textbf{5} (2015).

\bibitem{Palovics_2013}
R P\'{a}lovics, AA Bencz\'{u}r, Temporal influence over the last.fm social
  network in {\em Proceedings of the 2013 IEEE/ACM International Conference on
  Advances in Social Networks Analysis and Mining}, ASONAM '13.
\newblock (Association for Computing Machinery, New York, NY, USA), p.
  486–493 (2013).

\bibitem{triads}
S Wasserman, K Faust, {\em Social Network Analysis: Methods and Applications},
  Structural Analysis in the Social Sciences.
\newblock (Cambridge University Press), (1994).

\bibitem{Zhou_2018}
Z Zhou, K Xu, J Zhao, Homophily of music listening in online social networks of
  china.
\newblock {\em\protect\JournalTitle{Social Networks}} \textbf{55}, 160--169
  (2018).

\bibitem{Bisgin_2011}
H Bisgin, N Agarwal, X Xu, A study of homophily on social media.
\newblock {\em\protect\JournalTitle{World Wide Web}} \textbf{15}, 213--232
  (2011).

\bibitem{Salganik2006}
MJ Salganik, PS Dodds, DJ Watts, Experimental study of inequality and
  unpredictability in an artificial cultural market.
\newblock {\em\protect\JournalTitle{Science}} \textbf{311}, 854--856 (2006).

\bibitem{Salganik2008}
MJ Salganik, DJ Watts, Leading the herd astray: An experimental study of
  self-fulfilling prophecies in an artificial cultural market.
\newblock {\em\protect\JournalTitle{Social Psychology Quarterly}} \textbf{71},
  338--355 (2008).

\bibitem{Lynn2016}
FB Lynn, MH Walker, C Peterson, Is popular more likeable? choice status by
  intrinsic appeal in an experimental music market.
\newblock {\em\protect\JournalTitle{Social Psychology Quarterly}} \textbf{79},
  168--180 (2016).

\bibitem{Monechi_2017}
B Monechi, P Gravino, VDP Servedio, F Tria, V Loreto, Significance and
  popularity in music production.
\newblock {\em\protect\JournalTitle{Royal Society Open Science}} \textbf{4},
  170433 (2017).

\bibitem{svm}
Cw Hsu, Cc Chang, CJ Lin, A practical guide to support vector classification.
\newblock {\em\protect\JournalTitle{Journal of Machine Learning Research}}
  (2003).

\bibitem{svmpy}
scikit-learn developers, Svc in python sklearn
  (\url{https://scikit-learn.org/stable/modules/generated/sklearn.svm.SVC.html})
  (2022) Accessed: 2022-11-22.

\bibitem{Biau2016}
G Biau, E Scornet, A random forest guided tour.
\newblock {\em\protect\JournalTitle{{TEST}}} \textbf{25}, 197--227 (2016).

\bibitem{rfpy}
scikit-learn developers, Random forest in python sklearn
  (\url{https://scikit-learn.org/stable/modules/generated/sklearn.ensemble.RandomForestClassifier.html})
  (2022) Accessed: 2022-11-22.

\bibitem{Schapire2013}
RE Schapire, Explaining {AdaBoost} in {\em Empirical Inference}.
\newblock (Springer Berlin Heidelberg), pp. 37--52 (2013).

\bibitem{adbpy}
scikit-learn developers, Ada boost in python sklearn
  (\url{https://scikit-learn.org/stable/modules/ensemble.html#adaboost}) (2022)
  Accessed: 2022-11-22.

\bibitem{natekin2013gradient}
A Natekin, A Knoll, Gradient boosting machines, a tutorial.
\newblock {\em\protect\JournalTitle{Frontiers in neurorobotics}} \textbf{7}, 21
  (2013).

\bibitem{gbpy}
scikit-learn developers, Gradient tree boosting in python sklearn
  (\url{https://scikit-learn.org/stable/modules/ensemble.html#gradient-tree-boosting})
  (2022) Accessed: 2022-11-22.

\bibitem{knn}
T Cover, P Hart, Nearest neighbor pattern classification.
\newblock {\em\protect\JournalTitle{IEEE Transactions on Information Theory}}
  \textbf{13}, 21--27 (1967).

\bibitem{knnpy}
scikit-learn developers, K neighbors classifier in python sklearn (\url{
  https://scikit-learn.org/stable/modules/generated/sklearn.neighbors.KNeighborsClassifier.html})
  (2022) Accessed: 2022-11-22.

\bibitem{Hastie2009}
T Hastie, R Tibshirani, J Friedman, {\em The Elements of Statistical Learning}.
\newblock (Springer New York), (2009).

\bibitem{mlppy}
scikit-learn developers, Multi-layer perceptron classifier (\url{
  https://scikit-learn.org/stable/modules/generated/sklearn.neural_network.MLPClassifier.html})
  (2022) Accessed: 2022-11-22.

\end{thebibliography}

\clearpage
\onecolumn

\section*{Supplementary information}
\subsection{SI 1: Data availability of last.fm}
\label{SI1}

\begin{figure}[H]
	\centering
	  \includegraphics[width=0.6\columnwidth]{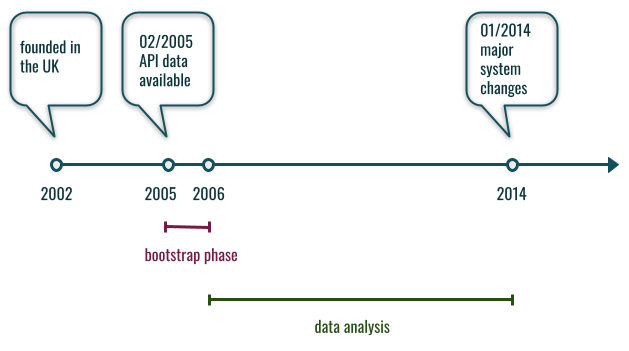}\\
		\caption{ 
		last.fm timeline and data availability. The company was founded in 2002 in the UK. User data is available on the API starting from February 2005. In 2014 there was a major change to the system - users were not able anymore to listen to music directly on last.fm but rather could connect their last.fm account to other streaming services such as Spotify. 
		}
	\label{fig:silfmt}
\end{figure}

Table \ref{tab3} shows the number of songs, listenings, albums, artists and users in the dataset that we collected from last.fm for this study. Users are further broken down into users for which the full friendship data i.e. the account names of all their friends are known and users for which we collected the full listen history. Figure \ref{fig:silfmt} shows the timeline of last.fm and marks the time periods for which data was fetched. The bootstrap phase was used to bootstrap the popularity of artists.

\begin{center}
\begin{table}[H]
\centering
\caption{Available data that was collected on the last.fm for the purpose of this study. Numbers are rounded down.}
\begin{tabular}{ c c }
 songs & 10,000,000  \\ 
 listenings & 300,000,000  \\  
albums & 200,000 \\
artists & 1,000,000 \\
users & 2,500,000 \\
with friendship data & 100,000 \\
with listen history & 17,500 
\end{tabular}
\label{tab3}
\end{table}
\end{center}

\subsection{SI 2: prediction parameters}
\label{SI:predictionparams}

We test several social-network based parameters in our prediction model. For these we use two networks: the friendship network and the influencer network. The friendship network of {\em last.fm} consists of nodes that represent users and links that represent bidirectional friendships. The influencer network is a directed network where users are nodes, and a weighted link represents the strength of a user’s influence over another. On these two networks, we define the following user metrics:

(a) the PageRank of each user (node), computed on the friendship network \\
(b) the nearest neighbor degree of each user, computed on the friendship network \\
(c) the degree of each user, computed on the friendship network \\
(d) the clustering coefficient of each user, computed on the friendship network \\
(e) the PageRank of each user, computed on the influencer network \\
(f) the nearest neighbor degree of each user, computed on the influencer network \\
(g) the degree of each user, computed on the influencer network \\
(h) the clustering coefficient of each user, computed on the influencer network \\
(i-k) user influence with time windows of 6h, 12h and 24h. The influence score of each user is divided by the number of songs where a user influenced another user thus giving us the number of influencing events per song. This is equivalent to the out-degree of the user in the influencer network divided by the number of songs.\\
(l) homophily as defined here by the average cosine similarity of the music preference vectors between a user and their friends\\

Each of these user metrics is computed for the users that contribute the first 200 listenings to a song and averaged per song. The average is then the value of the predictive parameter for that song. 

The performance of these parameters is compared to the performance of commonly used parameters that form our baseline. Specifically, we test the following parameters:

(m) the time, $\Delta t_j$, it takes a song, $j$, to reach the first 200 listenings in last.fm, given by

\begin{equation}
    \Delta t_j = t_{j, 200} - t_{j, 1}
\end{equation}

where $t_{j, i}$ is the timestamp in seconds when song, $j$, was listened to for the $i$th time. 

(n) the average number of listenings, that songs of the same genre receive in our dataset. We define the genre of a song, as the value of the the user-defined artist-tag with the largest weight (100). \\
(o) the average time that passes between two listenings of the same user, measured in seconds. Again, we only look at the first 200 listenings here. Whenever the same user listens more than once to the song withing those first 200 listenings, we compute the time that has passed in between. We then average these timespans per song.\\
(p) the number of users among the first 200 listenings that listen to the song again within a timespan of at most one week. This is similar to (o), but instead of looking at the time that passes in between, we simply count how many users listen to the song more than once. \\
(q) the area, $A_j$,  under the curve if the cumulative listenings of a song, $j$, up to a number of 200 are plotted vs time. This is given by  
\begin{equation}
    A_j = \int_{t = t_{j,1}}^{t_{j, 200}} l_{j}(t) dt
\end{equation}

where $t_{j, i}$ is the timestamp in seconds when song, $j$, was listened to for the $i$th time and $l_{j}(t)$ is the total number of times song, $j$, has been listened to at time, $t$.

(r) the same area as in (q), but subtracted from the area under the diagonal. This is given by

\begin{equation}
    A_j' = \frac{(t_{j, 200} - t_{j,1}) (l_j(t=t_{j, 200})-l_j(t=t_{j, 1}))}{2}  - \int_{t = t_{j,1}}^{t_{j, 200}} l_{j}(t) dt
\end{equation}

where $t_{j, i}$ is the timestamp in seconds when song, $j$, was listened to for the $i$th time and $l_{j}(t)$ is the total number of times song, $j$, has been listened to at time, $t$.

(s) the same area as in (q), but divided by the length of the x-axis (total time passed between the first and the 200th listening). This is given by

\begin{equation}
    A_j'' = \frac{\int_{t = t_{j,1}}^{t_{j, 200}} l_{j}(t) dt}{t_{j, 200} - t_{j,1}}
\end{equation}

where $t_{j, i}$ is the timestamp in seconds when song, $j$, was listened to for the $i$th time and $l_{j}(t)$ is the total number of times song, $j$, has been listened to at time, $t$.

(t) previous popularity of the artist. For this, we take a snapshot in time when the new song is released and calculate the average number of listenings that songs of the same artist have received in the past. \\

Table \ref{tab:siparams} shows the correlation coefficients for the parameters listed above and the popularity of songs.

\begin{center}
\begin{table}[H]
\centering
\caption{Pearson correlation coefficients between prediction parameters and song popularity. The prediction parameters were computed on the first 200 listenings for each song. Song popularity was defined as the maximum number of times a song was listened to {\em last.fm} in its best month. All p-values are below \boldsymbol{$10^{-50}$}.
}
\begin{tabular}{ c c }
(a) pagerank (Friendship NW) & -0.054 \\
(b) nearest neighbor degree (Friendship NW) & -0.081 \\
(c) degree (Friendship NW) & 0.090 \\
(d) clustering coefficient (Friendship NW) & 0.069 \\
(e) pagerank (Influencer NW) & 0.110 \\
(f) nearest neighbor degree (Influencer NW) & 0.090 \\
(g) degree (Influencer NW) &  0.132\\
(h) clustering (Influencer NW) & 0.062 \\
(i) influence 6h & 0.159 \\
(j) influence 12h & 0.169 \\
(k) influence 24h & 0.174 \\
(l) homophily & 0.081 \\
 & \\
(m) time to reach 200 listenings & -0.201 \\
(n) genre average & 0.069 \\
(o) time between repeated listenings & -0.193 \\
(p) number of repeated listeners  & 0.039 \\
(q) area under the trend curve & -0.223 \\
(r) area under the trend curve (subtracted from diagonal) & -0.085 \\
(s) area under the trend curve (normalized) & -0.044 \\
(t) previous popularity of artist & 0.135 \\
\end{tabular}
\label{tab:siparams}
\end{table}
\end{center}



\subsection{SI 3: Influencers and followers}
\label{SI:influ}
\begin{figure}[H]
    \centering
    \includegraphics[width=0.58\columnwidth]{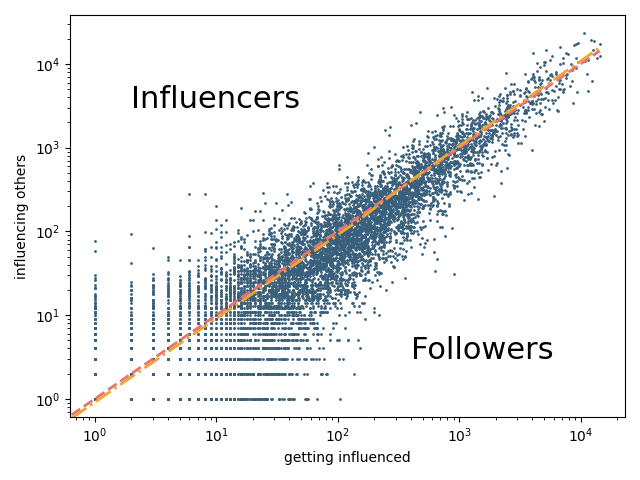}
    \caption{
	   Influencing others vs getting influenced by others. Each dot represents a user. The number of times a user was influenced into listening to a song is marked on the x-axis. The number of times a user influenced another user is marked on the y-axis. The red dashed line marks x=y. The yellow dashed and dotted line marks a power-law fit with $\alpha=1.009$. Users far above the red line tend to act as "influencers", while users far below tend to act as "followers".
	}
	\label{fig:influence_vs_influenced}
\end{figure}

In Fig. \ref{fig:influence_vs_influenced}, we report some heterogeneity of users in their influencing behavior. While some users tend to discover new songs early and (directly or indirectly) promote them in their friendship network --acting as ``influencers''--  others tend to follow these users and adopt their discoveries. We find that the higher the influence score of early song listeners, the quicker we can expect information about that song to spread in the friendship network. In the main text, we show how this information can improve machine learning models that predict the popularity of songs.

\subsection{SI 4: Homophily vs influence}
\label{SI:hom_vs_inf}

\begin{figure}[H]
    \centering
    \includegraphics[width=0.65\columnwidth]{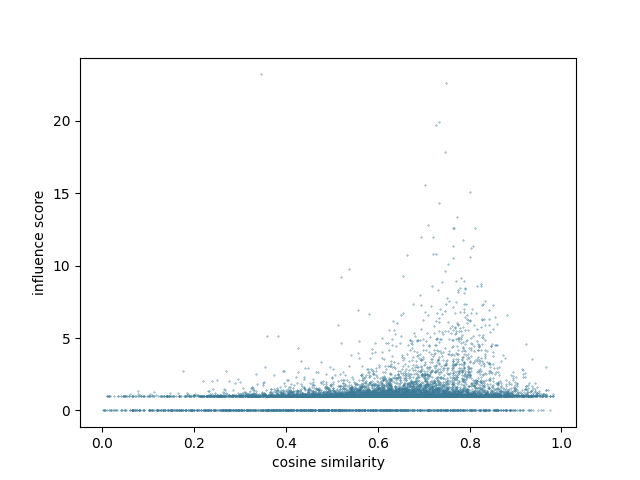}
    \caption{
	    User influence score vs average cosine similarity between a user and all of their friends. Each dot represents a user. The apparent gap between zero and one is due to to normalization of the influence score.
	}
	\label{fig:homophily_vs_influence}
\end{figure}

In figure \ref{fig:homophily_vs_influence} we show a scatterplot of how the cosine similarity of users relates to their influence score. 

\end{document}